\documentclass[a4paper,11pt]{article}
\pdfoutput=1 
\usepackage{subfigure}
\usepackage{jheppub} 
\usepackage[T1]{fontenc} 

\title{\boldmath Inspecting the power of parton showering the generated events by comparing the events generated in NLO+PS to NLO and NNLO without parton shower in $e^+e^-$ at 91.2 $GEV$}

\author[a]{Hessamoddin Kaveh,}
\author[b]{Mohammad Ebrahim Zomorrodian}

\affiliation[a]{MSc Student of Department of Science, Ferdowsi university of Mashad \\ Ferdowsi University of Mashhad, Azadi Sq., Mashhad, Khorasan Razavi, Iran.}
\affiliation[b]{Department of Science, Ferdowsi university of Mashad \\ Ferdowsi University of Mashhad, Azadi Sq., Mashhad, Khorasan Razavi, Iran.}	
\emailAdd{hessamoddin.kaveh@stu-mail.um.ac.ir}
\emailAdd{zomorrod@um.ac.ir}

\abstract{In this study we tried to emphasize the role parton shower plays in event generation, and in the physics of high energy event generation. We achieved this task by comparing the next-to-leading order and next-to-next-to-leading order results from EERAD3 with the next-to-leading order plus parton shower results from Vincia, and comparing both with real data.}

\keywords{$e^+e^-$, jet rate, jet resolution, parton shower, vincia, eerad3}

\begin{document}
\maketitle
\flushbottom

\section{Introduction}

In this study we tried to inspect the role played by parton showers in providing a better match to real data from the events generated by event generators, for a brief description of event generation and some mostly used event generators consult\cite{Buckley:2011ms}. To achieve this goal we generated events with Vincia\cite{Larkoski:2013yi} which is a plugin to Pythia\cite{Sjostrand:2007gs} to sample out how events with NLO+PS would look like; and we used EERAD3\cite{Ridder:2014wza} to generate fixed order NLO and NNLO events for comparison.

We have chosen jet rates and jet resolution variables for this comparison, as they contain a lot of physics so they will enable us to see the impact on physics as well. For jet finding we used Durham algorithm\cite{eekt}.Description of $e^+e^-$event variables are available in\cite{Dasgupta:2003iq}.

\section{Parton showers}
Due to the considerations we have for fixed-order event generation to work, we need to restrict ourselves to a region of phase-space where jets are hard and well-separated. In this regard we have omitted a lot of physics related to soft jets, jet substructure, etc\cite{Skands:2012ts}. These higher order effects can be implemented via a parton shower scheme.
Let us consider the cross section for the process $e^+e^- \longrightarrow q\bar{q}g$ which is the next-order process for $e^+e^- \longrightarrow q\bar{q}$.
\begin{equation}
\label{eqqg}
\frac{d\sigma_{q\bar{q}g}}{dcos\theta dz} \approx \sigma_{q\bar{q}}C_{F} 
\frac{\alpha_{s}}{2\pi} \frac{2}{sin^2\theta} \frac{1+(1-z)^2}{2}
\end{equation}
where $C_{F}=\frac{N_{c}^2-1}{2N_{c}}$ is a color factor\cite{Buckley:2011ms}.

In Eq.\eqref{eqqg} we see that the cross section for the next-order process is the cross section for the leading-order one, multiplied by something we interpret as the probability for gluon emission. We can further simplify it to a form more applicable to a Monte Carlo code. Considering the divergent regions and knowing their relation to collinearity of the gluon to the quark or the anti-quark we can express it as a sum on partons 
\begin{equation}
\label{generalqqg}
d\sigma_{q\bar{q}g} \approx \sigma_{q\bar{q}} \sum_{partons} C_{F} 
\frac{\alpha_{s}}{2\pi} \frac{d\theta^2}{\theta^2} dz \frac{1+(1-z)^2}{2}
\end{equation} 
and now $\theta$ is the angle between the gluon and the parton which emitted it.  

And it is also believed that the structure of Eq.\eqref{generalqqg} is mostly the same for any hard process like so. So for any hard process it can be achieved by introducing a color dependent variable $P_{ji}(z,\phi)d\phi$ instead of the $=C_{F}\frac{1+(1-z)^2}{2}$ part of the equation.\cite{Buckley:2011ms}. For the exact forms of cross sections see\cite{ellis2003qcd}.

By introducing an ordering variable and considering the probability of branching and non branching(Sudakov form factor) we can come up with an iterating scheme to attach additional partons to a hard process, one at a time\cite{Buckley:2011ms}. Adding branches with this method gives us a $2 \rightarrow n$ event from a $2 \rightarrow 2$ event.

In using parton showers we need to always be  aware of the double counting. As we might include higher orders we need to control the parton shower so it wont consider them for a second time.  Eliminating any double counted contribution from the matrix element matched to a parton shower, stops unphysical situations to happen.  

Vincia code uses antenna shower method which is described in\cite{Giele:2007di} and in essence is not very different from above.

\section{Jet finding}
Jets are one of the most amazing inventions of particle physicists; they give scientists the ability to interpret swarms of particles as different meaningful objects rather than raw data points.
To find jets in $e^+ e^-$ events, we need to use jet algorithms; the first widely used and a simple one of those is JADE algorithm. Proposed by JADE collaboration\cite{Bartel:1986ua}.For a history and description of most $e^+ e^-$ jet algorithms see\cite{Moretti:1998qx}. 
JADE algorithm roughly speaking is defining a distance variable $d_{ij} = \frac{2E_i E_j
(1-\cos\theta_{ij})}{Q^2}$, finding the minimum of it, defining a cut variable $y_{cut}$ and examining the minimum distances. Whether they are below the cut and combine into one or they are going to go into the iteration of the preceding process till there is no particle left uncombined. These entities which can not be combined any further  due to the cut variable, will be called jets\cite{Salam:2009jx}.
The JADE algorithm is infrared safe and collinear safe, however, because of the $E_i E_j$ part of the distance definition, it has the tendency to combine two very soft particles moving opposite to one another into one entity\cite{Salam:2009jx}. This property of JADE algorithm is catastrophic for a study of an event with parton showers at work.

In order to avoid such catastrophic incidents, we can use another jet algorithm which is quite the same as JADE. This algorithm is only slightly different in the definition of the distance variable
\begin{equation}
\label{Durham}
d_{ij} = \frac{2min(E_i^2 E_j^2)
(1-\cos\theta_{ij})}{Q^2}
\end{equation}
$Q$ is the total energy in the event, $\theta_{ij}$ is the angle between particles $i$ and $j$ and $E$ corresponds to their energies, just like the distance definition for JADE algorithm.\cite{Salam:2009jx} This new algorithm is called Durham/$k_t$ algorithm\cite{eekt}.

The difference made here makes the algorithm compatible with the situations in which there are very soft particles. In these situations the numerator of \eqref{Durham}, the distance variable reduces to the squared transverse momentum of $i$ relative to $j$. This property of the Durham algorithm is what gives it the $k_t$ name and saves it from failing in softer parts of the pahse-space.\cite{Salam:2009jx}

To use these algorithms and many more the Fastjet package can be utilized\cite{Cacciari:2011ma}. This package also provides a very efficient and fast way of doing so as the name suggests.
 
\begin{figure}[tbp]
\subfigure[3-jet fraction]{
\includegraphics[scale=0.25]{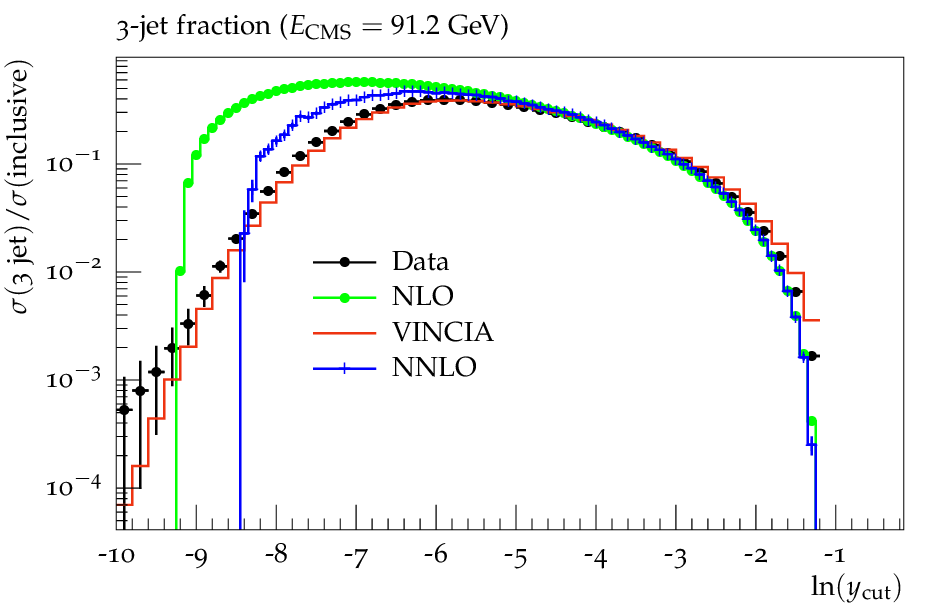}}
\subfigure[4-jet fraction]{
\includegraphics[scale=0.25]{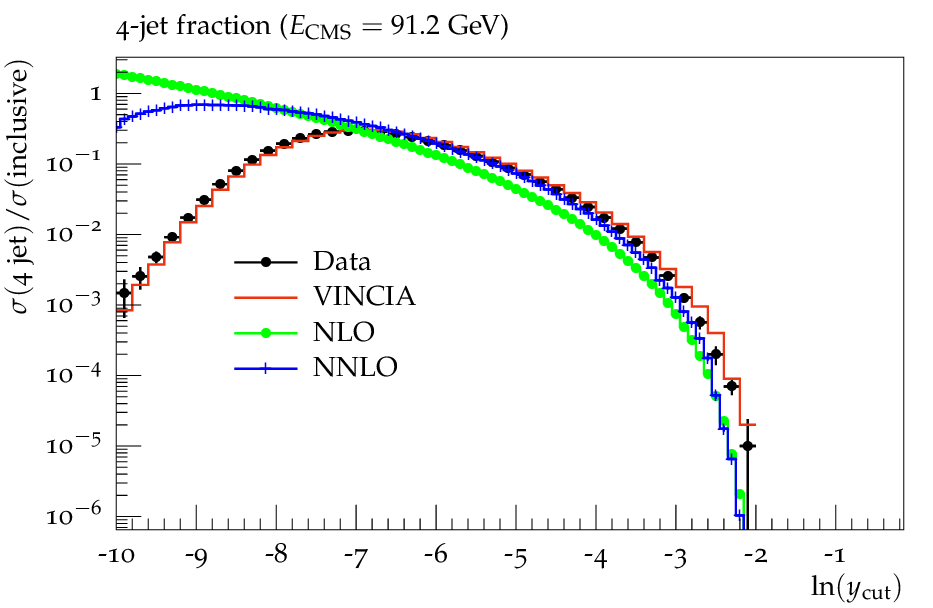}}
\subfigure[5-jet fraction]{
\includegraphics[scale=0.25]{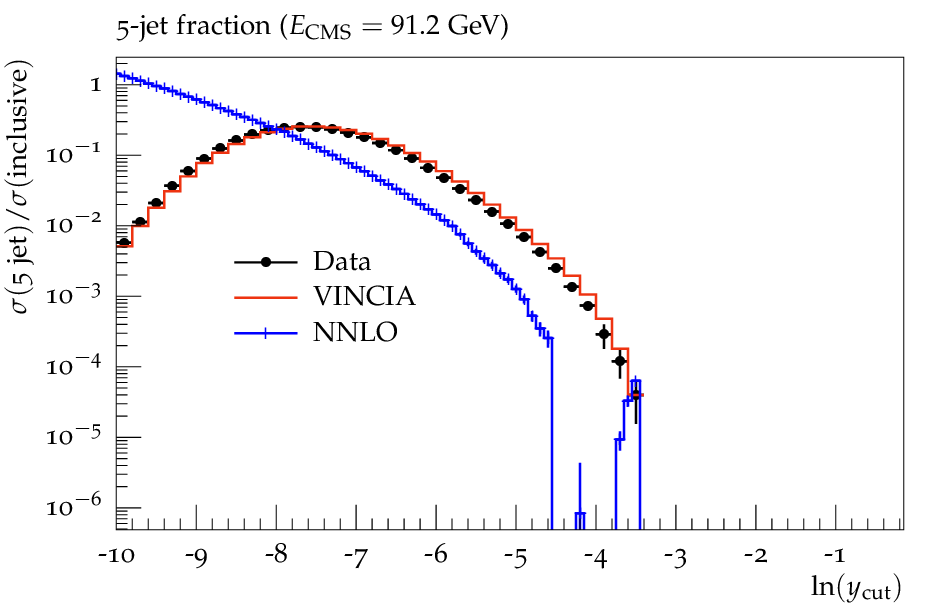}}
\caption{Jet rates at center of mass energy of 91.2 $GEV$ against $ln(y_{cut})$   }
\label{fig:1}
\end{figure}

\section{Results and conclusion}

The higher the energy of events, the more complicated they will become. This makes it very hard to mimic such processes. When we are dealing with complex events we need a far better input from theory to analyze or mimic data. To this avail a lot of scientific endeavors have dealt with more and more accurate calculations out of theories. Calculations of jet rates and jet resolution variables to higher orders have been one of these calculations. Calculations upto NNLO of event shape variables has been used in these\cite{Dissertori:2007xa,Khajooee:2014mva,Dissertori:2008cn} to estimate $\alpha_s$ and achieve NNLO corrections of it.

In this study we generated events with Vincia and EERAD3, compared our results with data from \cite{Heister:2003aj}.

To evaluate and test the data generated by Monte Carlo event generators Rivet \cite{Buckley:2010ar}is used.

If we consider figures \ref{fig:1}(a) and \ref{fig:1}(b) we can see that data at NLO and NNLO are present. Obviously NNLO is a better prediction than NLO as it is a better fit to data than the other. But when we compare the NNLO contribution against NLO+PS from Vincia we can deduce the fact that as $y_{cut}$ gets lower the result from Vincia provides a better fit to data than NNLO.

\begin{figure}[tbp]

\subfigure[$y_{23} $]{
\includegraphics[scale=0.25]{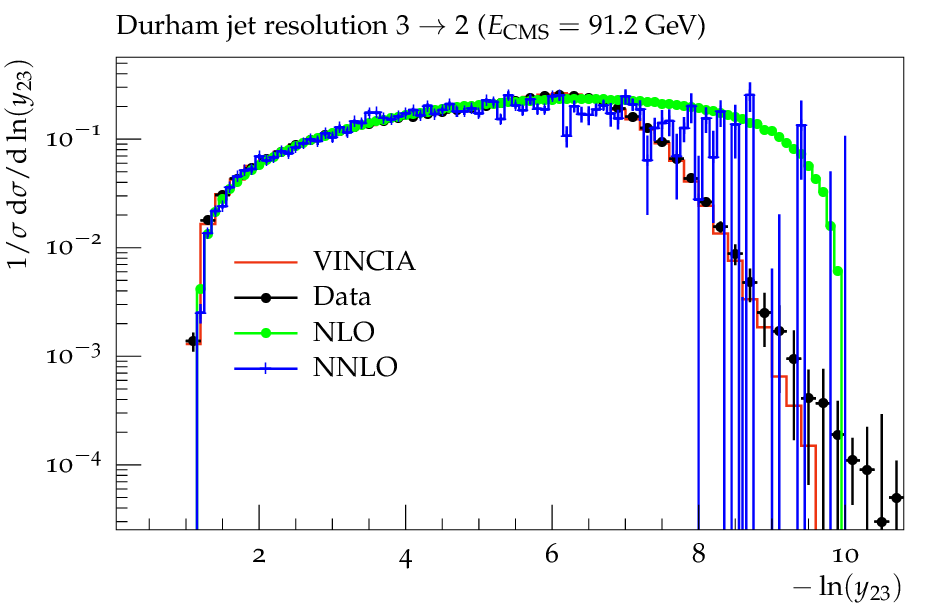}}
\subfigure[$y_{34}$]{
\includegraphics[scale=0.25]{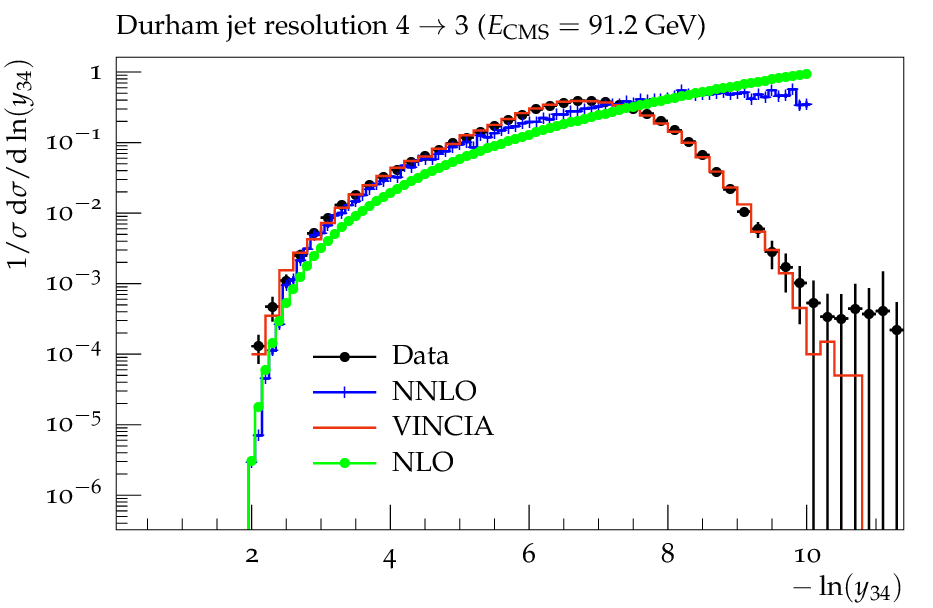}}
\subfigure[$y_{45}$]{
\includegraphics[scale=0.25]{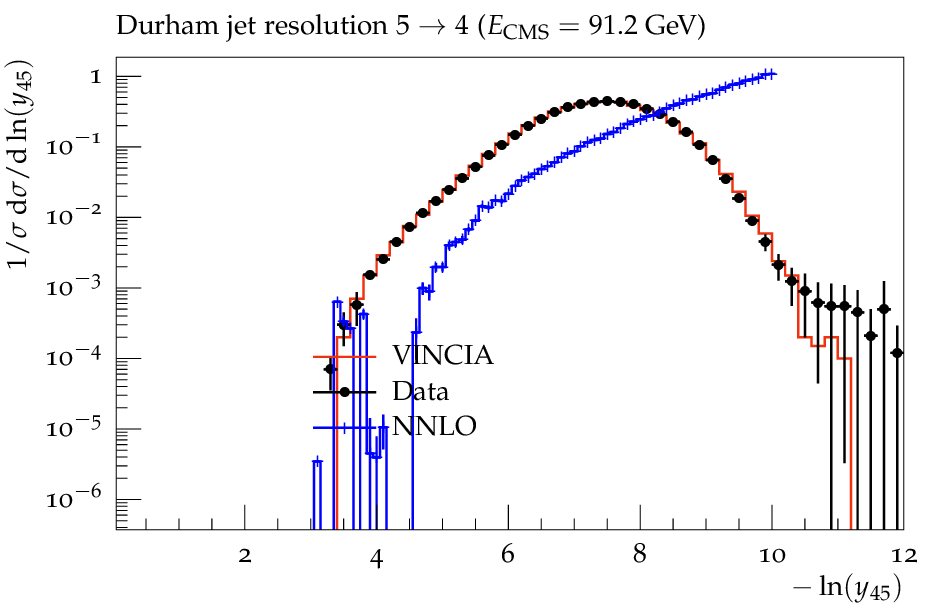}}
\caption{Jet resolution parameter at center of mass energy of 91.2 $GEV$ against -$ln(y)$   }
\label{fig:2}
\end{figure}

To understand why we get to a result like so we need to take a look at figures \ref{fig:2}(a) and \ref{fig:2}(b). These figures show the distance variable, also called the jet resolution variable.  What we can see here is the fact that when we get closer to the region  mapped out by lower $y_{cut}$s we are getting more and more into a region of phase-space which collinear and soft particles are dominating. We stated above that this region is not very well established in fixed order calculations, this intern causes the error in NNLO calculations,  shown in the figures utilizing error bars, are getting larger and larger approaching the lower $y_{cut}$ values. 

Now if we look at figures \ref{fig:1}(c) and \ref{fig:2}(c) which corresponds to the effects coming from the higher order calculations only. Although we have expected to see that NNLO is the better fit to data, we can see that a lot of contributions which came from parton showers exists even when the jets are hard and the region of phase-space is more filled with hard particles.We can see here that Vincia generated events, which we considered as NLO+PS accurate, certainly are providing the better fit to data.

As we discussed thoroughly above we can conclude that contributions from parton showers are inseparable from events. They contain a physics contribution, that a higher order calculation does not provide.

\end{document}